\documentstyle[prb,twocolumn,aps,psfig]{revtex}
\title{Quasiperiodicity, bistability and chaos in the Landau-Lifshitz equation}
\author{Luis Fern\'andez \'Alvarez$^1$, Oscar Pla$^1$, and Oksana 
Chubykalo$^{1,2}$}
\address{
$^1$Instituto de Ciencia de Materiales, Consejo Superior de Investigaciones
Cient{\'\i}ficas, Cantoblanco, E-28049 Madrid, Spain. 
}
\address{$^2$IBM Almaden Research Center, 650 Harry Rd., San Jose, CA 95120}

\date{\today}
\begin{document}
\draft
\twocolumn[\hsize\textwidth\columnwidth\hsize\csname@twocolumnfalse%
\endcsname

\maketitle

\begin{abstract}
The dynamics of an individual  magnetic moment is studied        
through the Landau-Lifshitz
equation with a periodic driving in the direction perpendicular to the
applied field. For fields lower than the anisotropy field and small
values of the perturbation amplitude we have
observed  the magnetic
moment bistability. At
intermediate values we have found quasiperiodic bands alternating with periodic
motion. At even larger values a chaotic regime is found. When the applied
field is larger than the anisotropy one, the behavior is periodic with
quasiperiodic regions. Those appear periodically in the amplitude of the
oscillating field. Also, even for low values of the driving force, the moment
is not parallel to the applied field.
\end{abstract}

]
\narrowtext

Traditionally, the study of the dynamics governed by the
Landau-Lifshitz equation is related to the ferromagnetic resonance
problems \cite{ferro}. Recently, the spin dynamics has become also
important in other
physical phenomena relevant to technological
applications, such as, e.g., magnetic recording processes
\cite{Bertram,Dieter} due to
a continuous increase of the magnetic recording
density together with the writing frequency\cite{Dieter,Stinnett} .
The dynamical micromagnetic calculations have
provided a useful tool in studying such important media
characteristic as dynamical coercitivity \cite{Chantrell}. 
 In spite of the fact that the Landau-Lifshitz 
equation is widely used in micromagnetic calculations, up to our
knowledge, no systematic study of its dynamics exists in the
literature. Let us recall here that this equation is nonlinear,
and in some regime one may expect a highly complicated dynamics
similar to one arising for an externally driven pendulum.
As an example, we can mention the nonlinear stochastic resonance
behavior of an individual magnetic moment \cite{Stochastic}. The
purpose of this paper is to present a systematic study of nonlinear
dynamics governed by the  Landau-Lifshitz equation, including
its bifurcation diagram and
stability properties. This would be very useful when analyzing results
obtained from large simulations of coupled Landau-Lifshitz (LL) equations.
There, and for some values of the coupling parameters, the individual
characteristics of each magnetic moment may play an important role in 
the collective
behavior. Eventually,
when enough of these moments are coupled, a description in terms of
magnons would be possible~\cite{prep}

Also the subject of chaos in magnetic materials~\cite{Wigen} is not new.
It has been studied in YIG, both experimental and theoretically, through spin
wave descriptions~\cite{Suhl,exp}. In different driving regime, the
chaotic behavior can arise in magnetostrictive wires and ribbons due
to the magnetoelastic coupling \cite{Hernando}.

In the original form (Landau-Lifshitz-Gilbert~\cite{Gilbert}) the equation may
be written as:
\begin{equation}
{\bf M}_t=-g\left({\bf M}\times{\bf H}_{\rm eff}\right)+\frac{\eta}{M_0}
\left({\bf M}\times{\bf M}_t\right),
\end{equation}
or in the more practical form (Landau-Lifshitz~\cite{LL})
\begin{equation}
\frac{1+\eta^2}{g}{\bf M}_t=-\left({\bf M}\times{\bf H}_{\rm eff}\right)
-\frac{\eta}{M_0}\left({\bf M}\times\left({\bf M}\times
{\bf H}_{\rm eff}\right)\right).
\end{equation}
Where $g$ is the local giromagnetic factor, $\eta$ is the damping, $M_0$ the
saturation magnetization, ${\bf M}$ is the tridimensional local continuous
magnetization, whose module is conserved (${\bf M}\cdot{\bf M} = M_0$), and
${\bf H}_{\rm eff}$ the effective field:
\begin{equation}
{\bf H}_{\rm eff}= {\bf H}_{\rm ext}+\beta{\bf n}\left({\bf n}\cdot{\bf M}
\right)+\alpha\nabla^2{\bf M}+{\bf H}_{\rm d}.
\end{equation}
Here ${\bf H}_{\rm ext}$ is the external magnetic field, $\beta$ is the anisotropy
coefficient, ${\bf n}$ is the unitary vector pointing in the anisotropy 
direction, $\alpha$ the coefficient of exchange, and, ${\bf H}_{\rm d}$ the
demagnetizing field that takes into account the dipolar interactions between
local moments in the system.

Here we will deal with only one local magnetic moment, so that 
we will not consider
exchange and dipolar interactions. This description could be relevant
to dynamics of magnetic microwires with a high anisotropy
\cite{Hernando}.

The external field will be decomposed
in two parts: ${\bf H}_n$, constant and parallel to the anisotropy direction,
and ${\bf h}={\bf h}_0\cdot\sin(\omega t)$, perpendicular to ${\bf n}$ and
oscillating in time with frequency $\omega$.

For practical purposes we rewrite
the equation in adimensional form and expanded notation:
\begin{mathletters}
\begin{eqnarray}
\kappa\dot{m_x}&=&-\left(m_y(h_z+m_z)-m_zh_y\right)-\eta\left(m_xm_z(h_z+
\right. \nonumber \\
&&\left. m_z)+m_xm_yh_y+(m_x^2-1)h_x\right), \\
\kappa\dot{m_y}&=&\left(m_x(h_z+m_z)-m_zh_x\right)-\eta\left(m_ym_z(h_z+
\right. \nonumber \\
&&\left. m_z)+
m_xm_yh_x+(m_y^2-1)h_y\right), \\
\kappa\dot{m_z}&=&-\left(m_xh_y-m_yh_x\right)-\eta\left(m_xm_zh_x+
\right. \nonumber \\
&&\left. m_ym_zh_y+(m_z^2-1)(h_z+m_z)\right),
\end{eqnarray}
\end{mathletters}
where ${\bf m}={\bf M}/M_0$, ${\bf h}= {\bf H}_{\rm ext}/(\beta M_0)$, 
$\kappa = 1+\eta^2$, and the dot represents the derivative with respect to
an adimensional time $\tau=g\beta M_0t$. We take the $z$ axis in the
direction of ${\bf n}$ and the $x$ axis in the direction of
${\bf h}_0$, which makes $h_y=0$.

These equations with appropriate initial conditions 
have been solved by using a fourth order Runge-Kutta scheme. We have observed
that the condition that $|{\bf m}|=1$ is fulfilled with a precision of
more than eight orders of magnitude ($10^{-8}$ in $1$), for even more than
$10^9$ integration steps. 

When looking for chaotic behavior in driven systems, the  usual quantities
to vary are the frequency and the amplitude of the perturbation. Sweeping in
frequency we have found that the interesting ones are those close to the
resonance frequency, $\omega_r = 1$, in our units, and of the order of several
GHz in real units. Thus possible perturbing signals are radio-wave sources. 
For simplicity in what follows will put the perturbing frequency fixed to the
resonance, and sweep in the amplitude, $h=(h_x^2+h_y^2)^{1/2}$ 

\begin{figure}
\mbox{\psfig{file=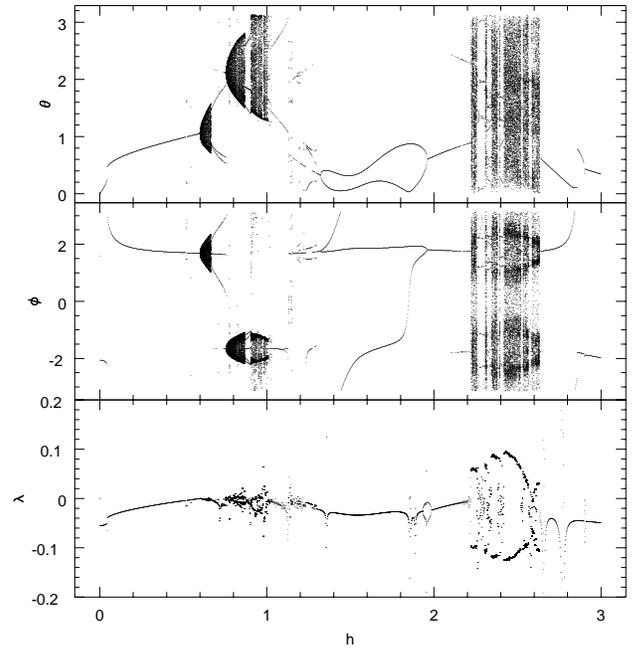,width=0.5\textwidth,angle=0}}
\caption{Bifurcation diagram for the $\theta$ and $\phi$ components of
the magnetization and Lyapunov exponents when $h_z=0.1$ and $\eta=0.05$.
\label{f:bif1}}
\end{figure}

A bifurcation diagram is shown in figure~\ref{f:bif1} for
$h_z=0.1$ and $\eta = 0.05$. Here only the components $m_\theta$ and
$m_\phi$  are drawn,
though in the numerical simulations we followed the equations
4a-c. The diagram  shows
the values of the components of ${\bf m}$, at time intervals multiple of 
$T=2\pi/\omega$ (Poincar\'e sections), with the value 
 $h$ shown in the $x$ axis of the figure.
There, various kinds of behaviors can be distinguished: 
periodic, quasiperiodic and chaotic motion. When there is only one
point for a given value of $h$, it represents a periodic motion with period
$T$; and when there is a continuum of points the behavior is quasiperiodic
or chaotic.
We will now try to describe the principal features of different critical
points shown in figure~\ref{f:bif1}. The changes in the diagram found when
the dissipation ($\eta$) is changed, are mainly quantitative (changes the value
of $h$ at which a given critical behavior is found). Also,the
 diagram shown has been
produced for the magnetic moment pointing in the direction of the external
field at $t=0$.

For small values of $h$ ($h\sim h_z$) a discontinuity that corresponds to a
folding bifurcation is found. This effect, consisting of two independent
limit circles, may be the experimental source for the observation of
hysteresis when changing the amplitude of the perturbation ($h$). When
increasing $h$ the jump in $\theta$ is the one shown in figure~\ref{f:bif1},
but if $h$ is decreased, the jump to smaller $\theta$ would occur at a lower
value of $h$. Experimentally, this phenomenon could manifest itself in
a bistable behavior of a magnetic microwire near the resonance
frequency.

\begin{figure}
\mbox{\psfig{file=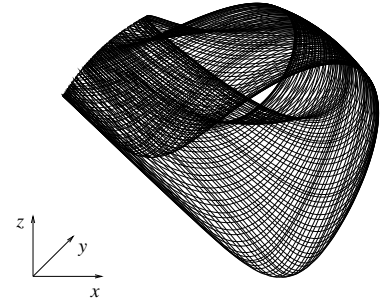,width=0.5\textwidth,angle=0}}
\caption{Quasiperiodic attractor at $h=0.64$ in the
 diagram of figure~\protect\ref{f:bif1}.
\label{f:muuu}}
\end{figure}

Two bifurcations, identified as torus have been observed at $h=0.60$ 
and $h=0.75$, leading to two regions of complex (but mainly quasiperiodic)
behavior: $h$ from 0.60 to 0.66, and 
from 0.75 to 0.87. The first one in the upper hemisphere, and the second one
in the lower. In the torus bifurcations the stable limit circles become
unstable and give rise to quasiperiodic motion on the surface of a torus. 
In figure~\ref{f:muuu} is shown the phase portrait, in coordinates $x$,
$y$, and $z$, of the quasiperiodic attractor at $h=0.64$.
There, the Poincar\'e section changes from just one single point to a
closed connected curve. 

\begin{figure}
\mbox{\psfig{file=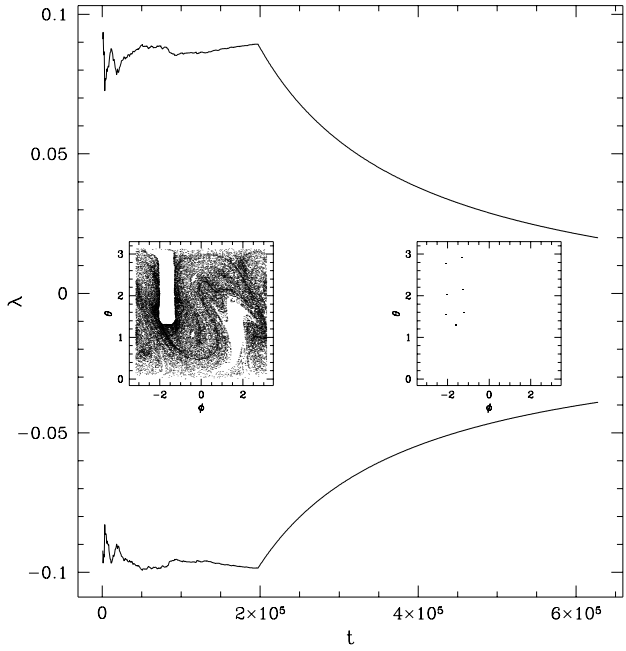,width=0.5\textwidth,angle=0}}
\caption{Time evolution for the Lyapunov exponents when $h=0.9785$ in the
diagram of figure~\protect\ref{f:bif1}. The left inset is the Poincar\'e
section taken from $t=0$ to $t=1.8\times 10^5$. The right one is taking
$t=4.5\times 10^5$ to $t=6.3\times 10^5$.\label{f:expo}}
\end{figure}

The region from 0.90 to 1.00 is a mixture of periodic and quasiperiodic
behavior, and even chaotic motion. The chaos in this region is characterized
by a chaotic attractor at $h\sim 0.9787$, which develops via a global
bifurcation of the type of chaotic transients. This means that the system
will evolve in the chaotic attractor for some time, and then, feeling the
periodic or quasiperiodic stable orbits, will leave it. This is illustrated
in figure~\ref{f:expo}, where the time evolution
of the Lyapunov exponents is shown. Initially both exponents converge, one
to a positive value, and the other to negative, as signature of chaos. But
at a given time the positive exponent initiates a decrease towards 0, or even
negative. The Poincar\'e sections for the initial and final time steps are
also shown. Initially the trajectories follow a chaotic map, but after
some time they eventually
fall in a period-seven orbit. The time spent in the chaotic behavior becomes
larger as the chaotic attractor is approached. 

\begin{figure}
\mbox{\psfig{file=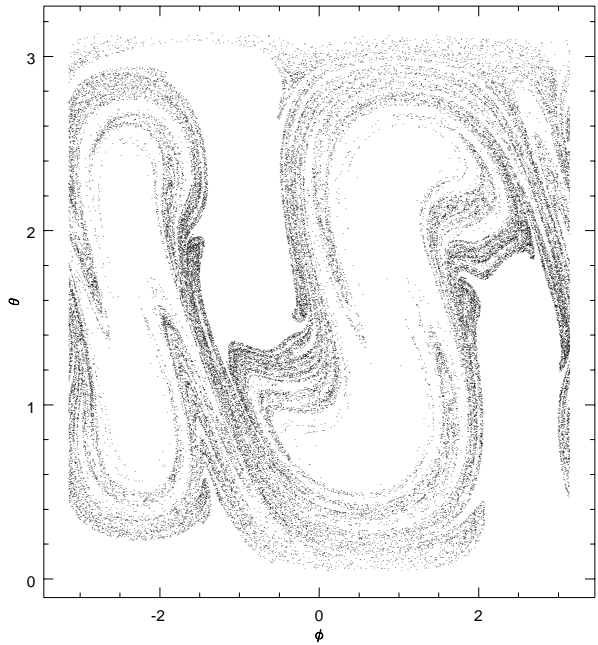,width=0.5\textwidth,angle=0}}
\caption{Chaotic attractor at $h=2.5$ in the
 diagram of figure~\protect\ref{f:bif1}.\label{f:mapa}}
\end{figure}

Next, there is a wide region of period doubling~\cite{bubble}, with some
higher period stripes. Finally, from 2.20 to 2.60, clear chaotic regions
(see the Lyapunov exponents in figure~\ref{f:bif1}) alternating with
quasiperiodicity are observed. 
In figure~\ref{f:mapa} we show the Poincar\'e section of the
chaotic attractor corresponding to $h$=2.50. In this case the route to chaos
is also that of chaotic transients.

If the initial state for the magnetization is in the direction opposite to the
external bias field, $h_z$, then, basically, the picture presented above holds.
Nevertheless the folding disappears, and the stable period-one orbit evolves
in the lower hemisphere. The same change of hemisphere happens for the
period doubling region. The two (upper and lower) torus bifurcations are also
preserved.

\begin{figure}
\mbox{\psfig{file=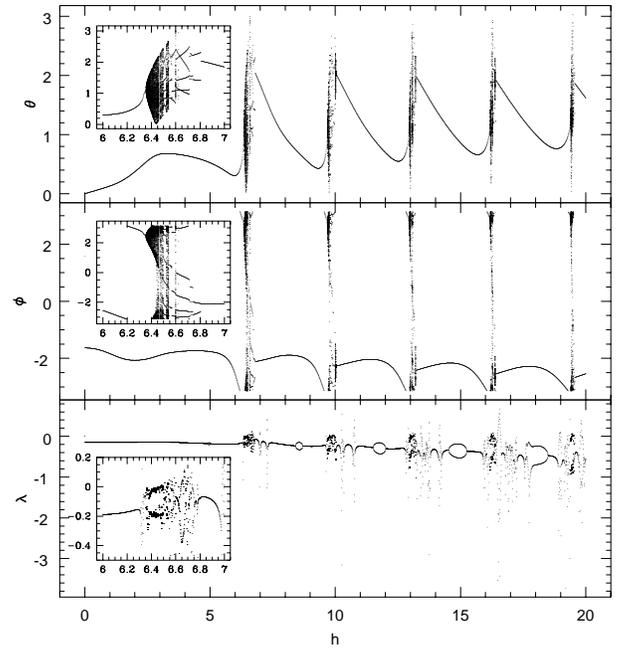,width=0.5\textwidth,angle=0}}
\caption{The same bifurcation as in fig.~\protect\ref{f:bif1}
for $h_z=2$. Insets show the magnification of the region of $h$ between
6 and 7.\label{f:bif2}}
\end{figure}

When the external applied bias, $h_z$, is larger than the anisotropy field
(larger than 1 in our units) the behavior is slightly different. The 
bifurcation diagram is shown in figure~\ref{f:bif2}. The same kind of
structures repeats at $h=3.25\times n$, where $n$ is an integer. A magnification
of those structures is shown in the inset, where it is seen that they consist
of a torus bifurcation and several periodic windows. The magnetization does
not remain at a value close to the saturation, but wanders over the whole
sphere.   

In conclusion, we have demonstrated that 
the dynamics governed by a driven Landau-Lifshitz
equation in certain parameter region can be very complicated. We have
observed manifestations of this behavior: the quasiperiodicity
alternating with periodic motion, the transient chaotic behavior and
finally a chaotic attractor. The experimental observation of these
predictions is limited to the materials which could have a well
defined single ferromagnetic resonance frequency, i.e. may be
described by a single magnetic moment. Nevertheless, some features of
this complicated behavior could be persistent in a coupled system 
~\cite{prep}. The estimation of the parameters show that even
in the case of a well-defined single magnetic moment behavior, the
experimental observation of the chaotic behavior would require a
powerful source of radio-waves. Nevertheless, a folding bifurcation
(bistable behavior) occurs for parameter values accessible by normal
radio-wave antennas. 

We acknowledge useful discussions with Jesus Gonz\'alez. L.F.A. is supported
by grant project PB96-0916 from CICYT.



\begin{references}
\bibitem{ferro}see,for example, L.R.Walker, in: {\sl Magnetism},
edited by G.T.Rado and H.Suhl, Academic Press, New York, 
pp. 337-348 (1963).
\bibitem{Bertram} see, for example, J.-G. Zhu, in: {\sl Magnetic
Recording Technology}, McGraw-HillNew York, pp.5.1-5.78 (1996).
\bibitem{Dieter} D.Weller and A.Moser, IEEE Trans. Magn., to be
published (1999).
\bibitem{Stinnett} S.M.Stinnett and W.D.Doyle, IEEE Trans.Magn. {\bf
34}, 1681 (1998).
\bibitem{Chantrell} R.W.Chantrell, J.D.Hannay, M.Wongsam, T.Schrefl,
H.-J.Richter, IEEE Trans. Magn. {\bf 34} 1839 (1998).
\bibitem{Stochastic} E.K.Sadykov and A.G.Isavin, Phys.Solid State {\bf
38} 1160 (1996).
\bibitem{Wigen}
See, for example P.E. Wigen (Ed.), {\sl Nonlinear Phenomena and Chaos in
Magnetic Materials}, World Scientific, Singapore, 1994.
\bibitem{Suhl}
X.Y. Zhang and H. Suhl, Phys. Rev. B{\bf 38}, 4893 (1988);
T.L. Carroll, F.J. Rachford, and L.M. Pecora, Phys. Rev. B{\bf 38}, 2938 
(1988); T.L. Carroll, L.M. Pecora, and F.J. Rachford, Phys. Rev. A{\bf 40}, 
377 (1889). 
\bibitem{exp}
D.W. Peterman, M. Ye, and P.E. Wigen, Phys. Rev. Lett. {\bf 74}, 1740 (1995);
S.M. Rezende, O.F. de Alcantara Bonfim, and F.M. de Aguiar, 
Phys. Rev. B, {\bf 33}, 5153 (1986);
F.M. de Aguiar, F.C.S. da Silva, and S.M. Rezende, Phys. Rev. E, {\bf 52},
2084 (1995).
\bibitem{Hernando} J.Vel\'{a}zquez, C.Garc\'{\i}a, M V\'{a}zquez and
A.Hernando, Phys. Rev. B {\bf 54}, 9903 (1996). 
\bibitem{prep} L. Fern\'andez, O. Pla, and O. Chubykalo, in preparation.
\bibitem{Gilbert}
T.L. Gilbert and J.K. Kelly, Proc of the Pitts. Conference on MMM, Pittsburgh,
Pa., 1955, p. 253 (unpublished); Phys. Rev. {\bf 100}, 1243 (1995).
\bibitem{LL} L. Landau and E. M. Lifshitz, Phys. Z. Soviet Union {\bf 8}, 153
(1935); L. Landau, {\sl Collected papers of Landau} (Pergamon, New York, 1965),
p. 110.
\bibitem{bubble} This region may be called also of period bubbling because
of the form the bifurcation diagram takes. It never proceeds further in
doubling the period, but returns to period halving.
\end{references}
\end{document}